\documentclass[twocolumn,showkeys,showpacs]{revtex4-1}
\usepackage[cp1251]{inputenc}
\usepackage{amsmath}
\usepackage{amsfonts}
\usepackage{amssymb}
\usepackage{graphicx}
\usepackage{textcomp}
\usepackage{color}
\begin{document}

\title{An improved kinetic Monte Carlo model for computational and analytical investigations of the magnetic properties of finite-size atomic chains}

\author{S.V. Kolesnikov}
\email{kolesnikov@physics.msu.ru}
\affiliation{Faculty of Physics, Lomonosov Moscow State University, Moscow 119991, Russian Federation}

\begin{abstract}
Two improved kMC models for investigations of the magnetic properties of finite-size atomic chains are presented. These models take the possible noncollinearity of magnetic moments into account. The spontaneous remagnetization of ferromagnetic Co chains on Pt(997) surface and antiferromagnetic Fe chains on $\text{Cu}_2\text{N/Cu(001)}$ surface is investigated in the framework of our models. The results are compared with the results of the simple kMC model. It is also shown that a single domain-wall approximation can be successfully used to estimation of the reversal time of the magnetization. Therefore, the improved kMC models can be used for analytical calculations as well as for computer simulations.
\end{abstract}	


\keywords{biatomic chains, magnetic properties, Heisenberg model, single domain-wall approximation}

\date{\today}
	
\maketitle

\section{Introduction}

Atomic chains and nanowires are very prospective in spintronics~\cite{ModPhisRev76.323}, quantum computing~\cite{Mermin_book}, quantum communications~\cite{PRL91.207901,EPL119.30001}, and other fields~\cite{RevModPhys91.041001} due to its unusual magnetic properties. The possibility of application of the atomic chains as bits of information appeared after the discovery of the giant magnetic anisotropy energy (MAE) of Co atoms in ferromagnetic atomic chains on the Pt(997) surface~\cite{Gambardella.Nature,PRL93.077203}. The possibility of creating and remagnetization of finite-sized antiferromagnetic chains was demonstrated for Fe atomic chains on Cu$_2$N/Cu(001) surface~\cite{science335.196,NatureNano10.40}.

Usually, the magnetic properties of the atomic chains can be satisfactory described in the the framework of Heisenberg model with uniaxial anisotropy. It is necessary to underline that quantum tunneling is the main switching mechanism at extremely low temperatures~\cite{JPCM27.455301}. However, the quantum nature of the atomic magnetic moments can be neglected at higher temperatures. In this case the classical Heisenberg model can be applied. If the external magnetic field is absent then the classical Heisenberg Hamiltonian can be written in the following form
\begin{equation}\label{eqHamiltonian}
H=-\sum_{i>j}J_{ij}\left({\bf s}_i\cdot{\bf s}_j\right)-K\sum_{i}\left({\bf s}_i\cdot{\bf e}\right)^2,
\end{equation}
where ${\bf s}_i$ and ${\bf e}$ are the unit vectors of the magnetic moments of the atoms and the easy axis of magnetization, respectively, $K$ is MAE, $J_{ij}=J(\delta_{i,j+1}+\delta_{i,j-1})$ is the exchange energy, $\delta_{ij}$ is Kronecker delta. For the ferromagnetic chains $J>0$ and for the antiferromagnetic chains $J<0$. The parameters of the Hamiltonian (\ref{eqHamiltonian}) can be found experimentally or calculated from the first principles by the means of density functional theory~\cite{DFT} or Korringa-Kohn-Rostoker-Green's function method~\cite{RPP74.096501}. Further investigation of the magnetic properties of atomic chains can be performed with either the solution of the Landau-Lifshitz-Gilbert equation~\cite{Landau,PRB93.161412R} or the Monte Carlo simulations~\cite{kMC_book}.

The simplest kinetic Monte Carlo (kMC) model for investigation of magnetic properties of atomic chains was proposed by Li and Liu~\cite{PhysRevB.73.174418}. This kMC model allows to calculate the critical temperature, the reversal time of the magnetization, and the coercive field of ferromagnetic chains~\cite{JMMM378.186,NJPhys11.063004,PRL96.217201,CPB24.097302,PSS57.1513}. It can be also applied for the investigation of antiferromagnetic chains~\cite{PRB93.035444,MPLB.31.1750142}. However, the simplest kMC model~\cite{PhysRevB.73.174418} assumes that (i) the directions of all magnetic moments are collinear to the easy axis of magnetization and (ii) rotation of the magnetic moment does not influence on the directions of other ones. These assumptions are very rude, because the metastable states of ferromagnetic or antiferromagnetic chains can be noncollinear~\cite{PRB57.5923,PRB81.054440,PRB92.024414}.

The main goal of our paper is introducing the improved kMC models which take the noncollinearity of magnetic moments into account. In Section~\ref{Methods} we describe two improved kMC models. The applicability of these kMC models to real systems (Co on Pt(997) and Fe on $\text{Cu}_2\text{N/Cu(001)}$) are shown in Section~\ref{RandD}. Section~\ref{Conclusion} concludes the paper. Some additional remarks about the calculation of diffusion barriers are presented in Appendix.

\section{Methods}\label{Methods}

This is the most important Section of the paper. Here we present two novel kMC models which significantly improve computational accuracy. These models take the noncollinearity of magnetic moments into account. In Section 3 the results of the presented models will be compared with the analogous results of the simple kMC model~\cite{PhysRevB.73.174418}. So, first of all we remember the simple kMC model. At the end of this Section the analytical method~\cite{Kolesnikov1,Kolesnikov2,Kolesnikov3,Kolesnikov4} which allows to estimate the reversal time of magnetization of atomic chains is shortly discussed.

\subsection{A simple kMC model}

The simplest kMC model for investigation of magnetic properties of atomic chains is the model suggested by Li and Liu~\cite{PhysRevB.73.174418}. It is assumed that all of the magnetic moments are directed either parallel or antiparallel to the easy axis of magnetization $\left({\bf s}_i\cdot {\bf e}\right)=\pm1$. Each magnetic moment can be only in two states: it is directed ``up'' if $\left({\bf s}_i\cdot{\bf e}\right)=1$, or ``down'' if $\left({\bf s}_i\cdot{\bf e}\right)=-1$. The transition of $i$th magnetic moment from one state to another is its rotation in the plain at the condition that all other magnetic moments are still directed either up or down. In this case the transition rates $\nu_i$ can be easily calculated analytically. If $2K>|h_i|$ then
\begin{equation}\label{eqRate1}
\nu_i=\nu_0\exp\left(-\frac{\left(2K+h_i\right)^2}{4Kk_\text{B}T}\right),
\end{equation}
where $k_\text{B}$ is the Boltzmann constant, $T$ is the temperature, and $\nu_0=10^9$~Hz~\cite{Gambardella.Nature} is the frequency prefactor, $h_i=\sum_{j}J_{ij}({\bf s}_i\cdot{\bf s}_j)$.
If $2K\le|h_i|$, then there is no diffusion barrier between the initial and the final states. The transition rate $\nu_i$ can be calculated~\cite{Glauber1.1703954}, as
\begin{equation}\label{eqRate2}
\nu_i=\nu_0\frac{\exp(-2h_i/k_\text{B}T)}{1+\exp(-2h_i/k_\text{B}T)}.
\end{equation}

\subsection{An improved kMC model I}

The simple kMC model~\cite{PhysRevB.73.174418} assumes that the directions of all magnetic moments (except one) are frozen. This assumption is the main drawback of the model. In order to overcome this drawback we suggest that the magnetic moments can be noncollinear to each other and the easy axis of magnetization. Thus, we can not manifest that the $i$th magnetic moment is directed up or down. Instead of this the directions of all magnetic moments should be found as a result of minimisation of the magnetic energy of the chain. The minimisation of the magnetic energy can be realised as a relaxation of the directions of the magnetic moments. This process is analogous to finding of relaxed geometry of an atomic system by means of molecular statics. We also can calculate the diffusion barriers between the relaxed states by means of the geodesic nudged elastic band (GNEB) method~\cite{GNEB}. The symmetry of the Hamiltonian (\ref{eqHamiltonian}) allows to significantly simplify calculations by usage of XY-model. Important details of molecular statics and GNEB methods in the framework of XY-model are shortly discussed in Appendix.

The thickness of the domain wall can be estimated as $\delta N=\sqrt{J/2K}$ atoms~\cite{Lundau_book,Kolesnikov3}. We can distinguish two different cases: $J\ll K$ and $J\gtrsim K$. Let us discuss the first case. If $J\ll K$ then the thickness of the domain wall is neglectable ($\delta N\ll 1$). Thus, all magnetic moments are slightly collinear to the easy axis of magnetization $\left({\bf s}_i\cdot {\bf e}\right)\approx\pm1$. It means that the atomic chain has the same metastable states as in the framework of the simple kMC model~\cite{PhysRevB.73.174418}. In other words, we can still assume that only one magnetic moment flips at each kMC step. However, instead of the analytical equation (\ref{eqRate1}) we use the general rule of calculations of the transition rate
\begin{equation}\label{eqRate3}
\nu_i=\nu_0\exp\left(-\frac{E^D_i}{k_\text{B}T}\right),
\end{equation}
where $E^D_i$ is diffusion barrier calculated by means of GNEB method. This kMC model we will refer as the improved kMC model I. All diffusion barriers are calculated ``on the fly'' and saved in database. From this point of view the improved kMC model I is analogous to self-learning kMC models widely used for simulation of diffusion processes~\cite{JCP115.9657,PRB72.115401,SS612.48}.

Let as underline the main features of the improved kMC model I:
(i) there are the same metastable states as in the simple kMC model~\cite{PhysRevB.73.174418},
(ii) all diffusion barriers are calculated ``on the fly'' with GNEB method~\cite{GNEB}.

\subsection{An improved kMC model II\label{SecModII}}

Let us discuss the case $J\gtrsim K$. The domain wall has the thickness of several atoms. Thus, the straightforward rotations of single magnetic moments can lead to unstable states. As a consequence, the number of metastable states of the atomic chain is different to one in the framework of the simple kMC model~\cite{PhysRevB.73.174418}. Therefore, searching the metastable states becomes the important part of the kMC algorithm. There are exact methods of searching the metastable states of finite-size chains~\cite{PRB57.5923,PRB81.054440}. However these methods are hardly applicable to the atomic chains consisting of several tens of atoms because of their very high computational cost in the case of long chains.

Here, we present the following method of searching the metastable states of finite-size chain. First of all we need find all low energy metastable states of an infinite chain. There are four such states (see also Figure~\ref{fig3}): clockwise domain wall (CDW), anti-clockwise domain wall (ACDW), clockwise anti-domain wall (CADW), and anti-clockwise anti-domain wall (ACADW). All these states have the same energy. We will refer their as etalon states. Now, we can use the etalon states for searching of metastable states of the finite-size chain. This procedure consist of two steps: (i) constructing of the metastable state from the etalon states, (ii) relaxation of the atomic chain. Each metastable state can be labelled with the the number of etalon states and their positions. So, recognising of any metastable state can be performed by mapping with the etalon states.

At each kMC step one of the following events occurs: (i) formation or disappearance of the etalon states at the edge of the chain, (ii) formation or disappearance of pair of the etalon states (CDW-CADW or ACDW-ACADW), (iii) transition of the etalon state along the chain, and (iv) transition of the clockwise etalon state to anti-clockwise etalon state or vise versa. All diffusion barriers are calculated ``on the fly'' by means of GNEB method. This kMC model we will refer as the improved kMC model II.

Let as underline the main features of the improved kMC model II:
(i) searching of metastable states by mapping with etalon states,
(ii) existence of clockwise and anti-clockwise states,
(iii) all diffusion barriers are calculated ``on the fly'' with GNEB method~\cite{GNEB}.

\subsection{Analytical method}

The reversal time of the magnetization of the atomic chain can be easily calculate in the framework of a single domain-wall approximation~\cite{Kolesnikov1,Kolesnikov2,Kolesnikov3,Kolesnikov4}. The idea of this method is the following. The reversal time of the magnetization $\tau$ of the atomic chain can be calculate as the average time of the random walk of the domain wall. In the simplest case the random walk of the domain wall is characterized by only three rates: (i) the rate of formation of the domain wall $\nu_1$, (ii) the rate of the domain wall disappearance $\nu_2$, and (iii) the rate of motion of the domain wall along the chain $\nu_3$. To calculate the average time of the random walk of the domain wall the mean rate method can be employed~\cite{JCP132.134104,PMA76.565}.

In the simplest case the reversal time of magnetization of ferromagnetic or antiferromagnetic single-atomic chain can be obtained as
\begin{multline}\label{eqTau}
\tau=\frac{1}{na}\left\{\frac{a}{\nu_3}\left(\frac{N-1}{2}\right)\left[N-\frac{2(1-2a)}{1-a}\right]\right.+\\
+\left.\frac{1}{\nu_1}\left[N(1-a)-2(1-2a)\right]\right\},
\end{multline}
where $a=\nu_3/(\nu_2+\nu_3)$, $n=2$, $N$ is number of atoms in the chain. Equation (\ref{eqTau}) has been derived in the framework of the simple kMC model~\cite{PhysRevB.73.174418} and can be used without any changes in the framework of the improved kMC model I. Here $n$ is the number of states which can form at the edges of the chain: domain wall from the one side of the chain and anti-domain wall from another side of the chain.

In order to use equation (\ref{eqTau}) in the case of the improved kMC model II the following two changes should be applied. First, there are four states (CDW, ACDW, CADW and ACADW) which can form at the edges of the chain. Thus, $n=4$. Second, the etalon states have nonvanishing thickness. Thus, the number $N$ should be replaced by $N_\text{eff}\le N$, where $N_\text{eff}-1$ is a number of possible metastable positions of CDW (or another etalon state) in the atomic chain.

\section{Results and Discussions}\label{RandD}

In order to illustrate applicability of the proposed kMC models to real atomic chains we shortly discuss two systems: Fe chains on $\text{Cu}_2\text{N/Cu(001)}$ surface and Co chains on Pt(997) surface.

\subsection{Antiferromagnetic Fe chains on $\text{Cu}_2\text{N/Cu(001)}$ surface}

\begin{figure}[htb]
\begin{center}
\includegraphics[width=0.95\linewidth]{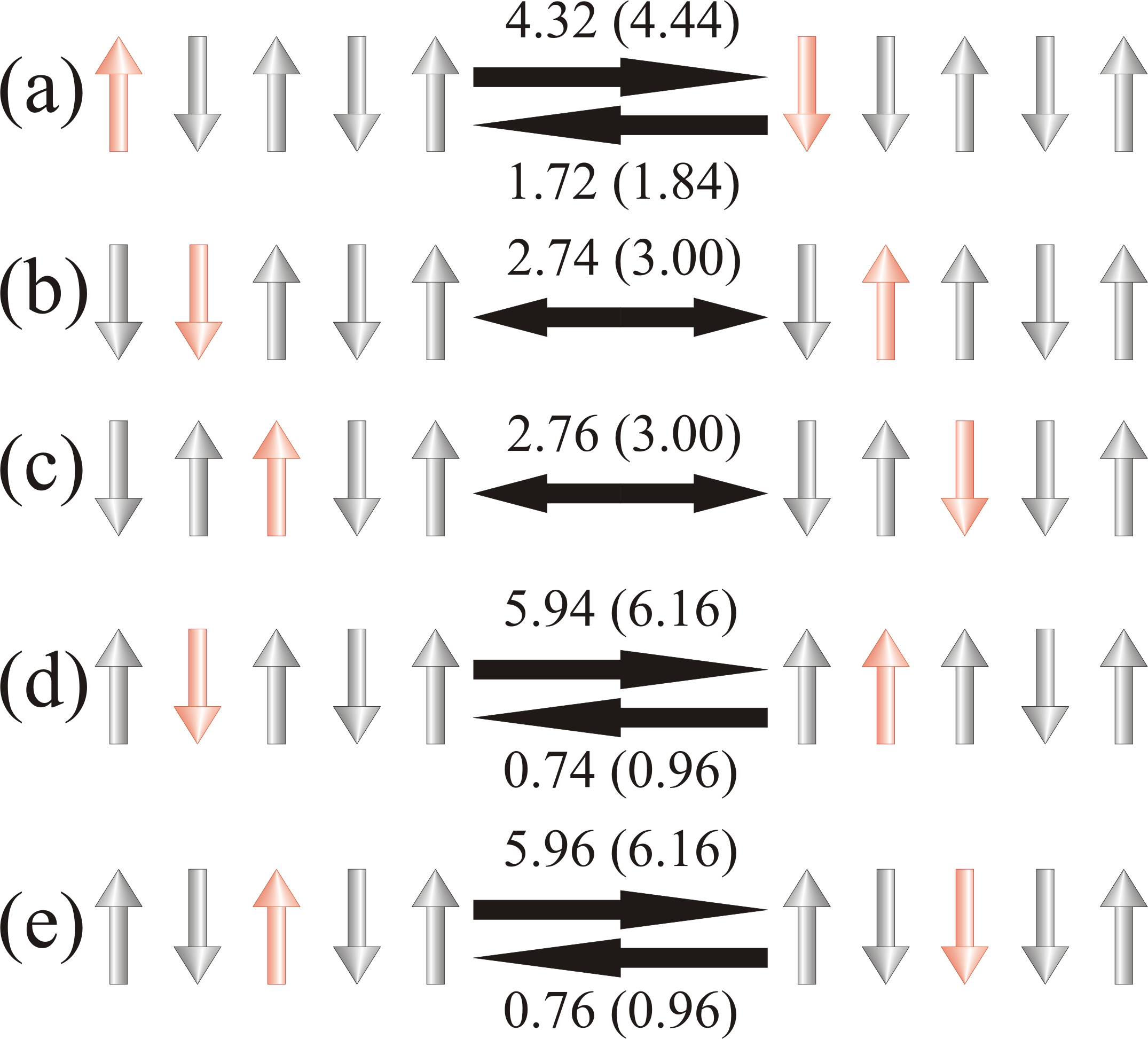}
\caption{\label{fig1} Diffusion barriers of the rotations of magnetic moments in the case of Fe chain consisting of $N=5$ atoms on $\text{Cu}_2\text{N/Cu(001)}$ surface: (a) formation of the domain wall at the edge, (b,c) motion of the domain wall along the chain, (d,e) formation of the pair of domain walls. All values are given in meV. The nonrelaxed values (simple kMC model~\cite{PhysRevB.73.174418}) are given in brackets.
}
\end{center}
\end{figure}

According to the experimental study~\cite{science335.196}, the exchange energies of Fe atoms are $J=1.3\pm0.1$~meV. In three-atomic chain MAE varies from $2.1\pm0.1$~meV for the edge atoms to $3.6\pm0.1$~meV for the central atom~\cite{NatureNano10.40}. For the numerical estimates the following parameters of the Hamiltonian (\ref{eqHamiltonian}) $J=1.3$~meV, $K=3.0$~meV are chosen. The ratio $K/J\approx 2.3$ and the rotations of single magnetic moments always lead to new metastable state. The examples of such events in the case of very short Fe chain consisting of $N=5$ atoms are shown in Fig.~\ref{fig1}. One can see that the taking the relaxation effects into account leads to decreasing the diffusion barriers. Moreover, the diffusion barriers at the edges of the chain are less then the analogous barriers at the middle of the chain. However, the thickness of domain walls is neglectable, all magnetic moments in metastable states are approximately parallel to the easy axis of magnetization, and the number of metastable states is the same as in the simple kMC model. Therefore, the improved kMC model I can be applied for this system.

\begin{figure}[htb]
\begin{center}
\includegraphics[width=0.95\linewidth]{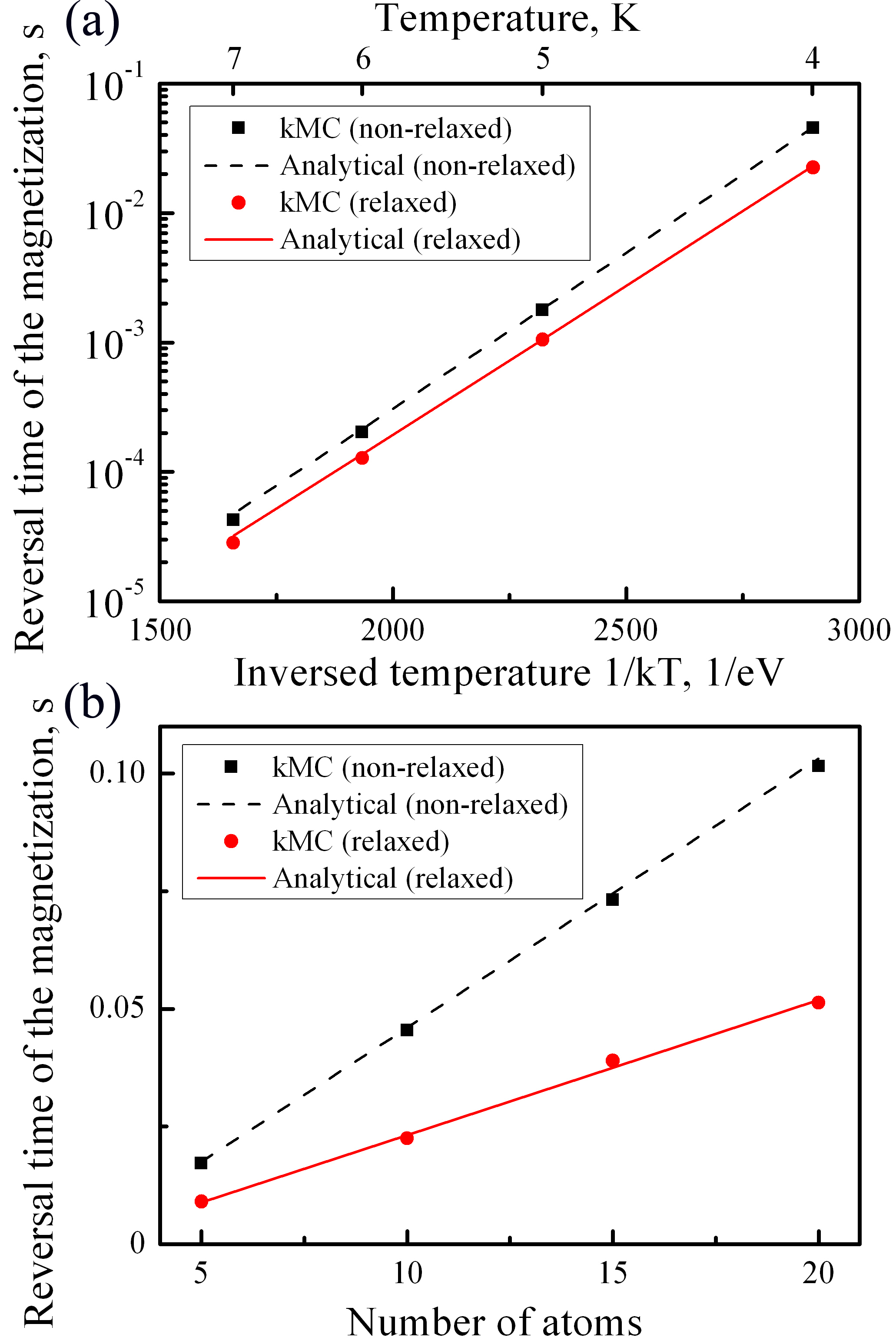}
\caption{\label{fig2} The dependencies of the reversal time of the magnetization the Fe chain on $\text{Cu}_2\text{N/Cu(001)}$ surface in the framework the simple kMC model~\cite{PhysRevB.73.174418} and the improved kMC model~I. (a) The temperature dependence of the chain consisting of $N=10$ atoms. (b) The dependence on the length of the chain at $T=4$~K. The parameters of the Heisenberg Hamiltonian are the following:
$J=1.3$~meV and $K=3.0$~meV.
}
\end{center}
\end{figure}

Figure~\ref{fig2} shows the dependencies of the reversal time of the magnetization the Fe chain on $\text{Cu}_2\text{N/Cu(001)}$ surface. The results of kMC simulation are averaged over 1000 remagnetisations and shown with points. In order to estimate an influence of the relaxation effect on the reversal time of the magnetization we compare the results of the simple kMC model (black points) and the improved kMC model I (red points). Everyone can see that the relaxation effect leads to decrease of the reversal time of the magnetization by the factor $\eta=\tau/\tau_\text{relaxed}\approx2-3$ at $T=4-7$~K (Fig~\ref{fig2}a). The factor $\eta$ is almost independent on the length of the chain. Indeed, at the current parameters of the Hamiltonian the following inequalities are satisfied: $\nu_1\ll \nu_3\ll \nu_2$. Therefore, the equation (\ref{eqTau}) can be simplified as
\begin{equation}\label{eqTau2}
\tau\approx\frac{N-2}{2}\frac{\nu_2}{\nu_1\nu_3}.
\end{equation}
At this limit the the reversal time of the magnetization linearly depends on the number of atoms $N$ (Fig~\ref{fig2}b), and the factor $\eta$ does not depend on $N$.

The analytical results obtained with equation (\ref{eqTau}) are presented in Fig.~\ref{fig2} with solid and dashed lines. In the case of the simple kMC model the equation (\ref{eqRate1}) are used for calculation of the transition rates $\nu_i$, $i=1,2,3$. In the case of the improved kMC model I we use the equation (\ref{eqRate3}) with the diffusion barriers shown in Fig.~\ref{fig1}: $E^D_1=4.32$~meV, $E^D_2=1.72$~meV, $E^D_3=2.76$~meV. One can see that our analytical approach
gives the same results as the kMC simulations.

\subsection{Ferromagnetic Co chains on Pt(997) surface}

According to the experimental paper~\cite{Gambardella.Nature}, the exchange energies of Co atoms are $J\approx7.5$~meV and MAE is $2.0\pm0.2$~meV. For the numerical estimates we choose the following parameters of the Hamiltonian (\ref{eqHamiltonian}): $J=7.5$~meV and $K=2.0$~meV. The ratio $K/J\approx 0.27$, and the thickness of the domain wall is not neglectable ($\delta N=\sqrt{J/2K}\approx1.37$). In this case the improved kMC model II can be applied, and the metastable states can be found as it was described in Section~\ref{SecModII}. If $N<10$ then there are not metastable states. In this case all magnetic moments flip simultaneously (superparamagnetic regime). If $10\le N<22$ then only the single etalon states (CDW, ACDW, CADW and ACADW) can be metastable. If $N=10$ the metastable etalon states can be located only in the middle of the chain. If $N=22$ then the pair of states CDW-ACADW and CDW-CADW can be metastable. If $N>22$ then all possible pairs of etalon states can be metastable.

\begin{figure}[htb]
\begin{center}
\includegraphics[width=0.7\linewidth]{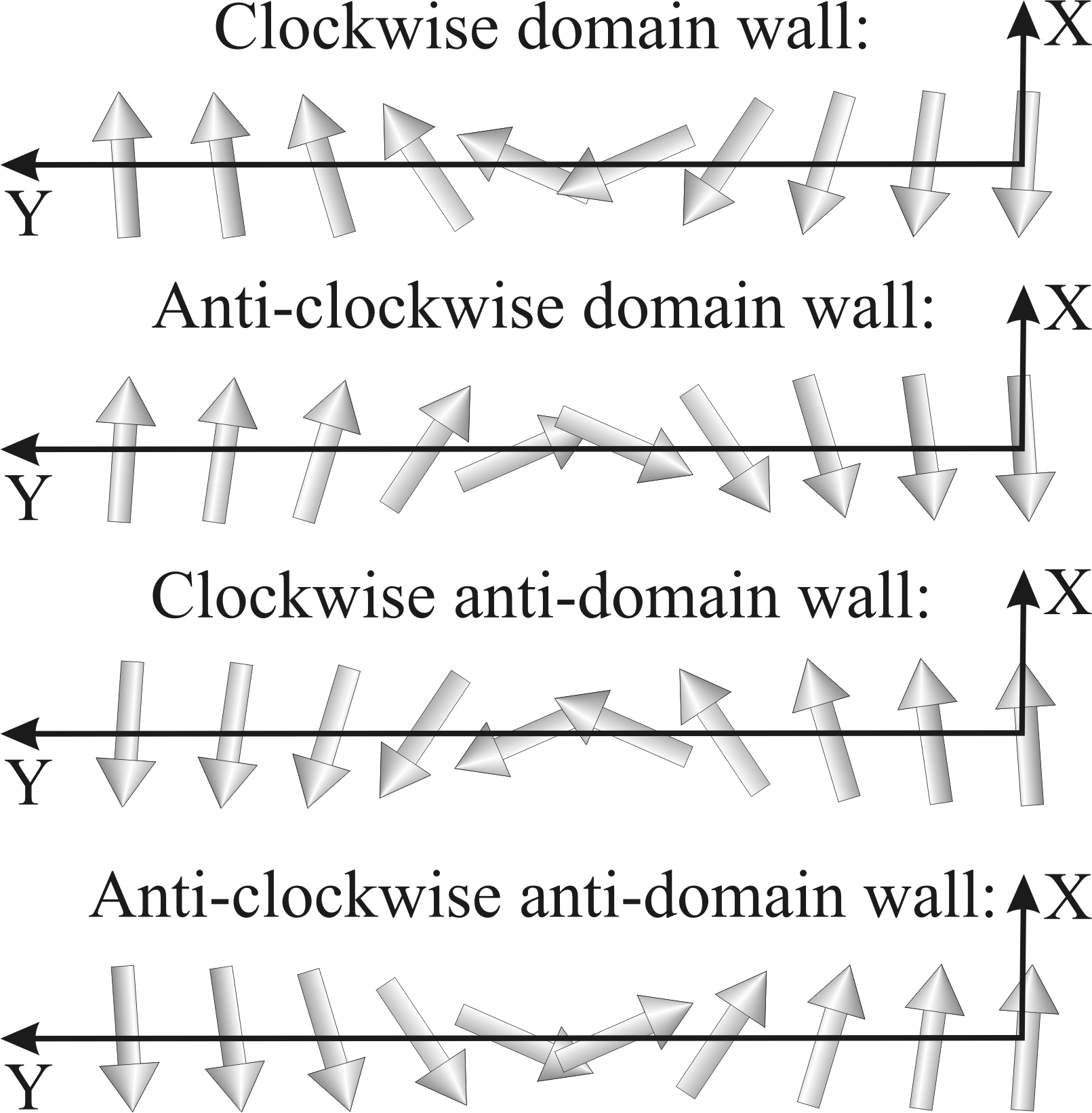}
\caption{\label{fig3} Possible one-domain wall states of the Co chain on Pt(997) surface in the framework of the improved kMC model~II. The central ten atoms of the long chain are shown.
}
\end{center}
\end{figure}

For searching of metastable states we use the etalon states located in the middle of the Co chain consisting of $N=20$ atoms. The central ten magnetic moments are shown in Fig.~\ref{fig3}. Magnetic moments rotates in XY-plane. It is also assumed that the atoms placed along the Y axis. The etalon state can not be metastable if it is located very close to the edge of the chain because it has nonvanishing thickness. If $N\ge12$ then the first metastable etalon state is located between the 6th and 7th atoms of the chain. In other words, five possible position from each side of the chain is unstable. Thus, the number $N$ in equation (\ref{eqTau}) should be replaced by $N_\text{eff}=N-10$.

\begin{figure}[htb]
\begin{center}
\includegraphics[width=0.95\linewidth]{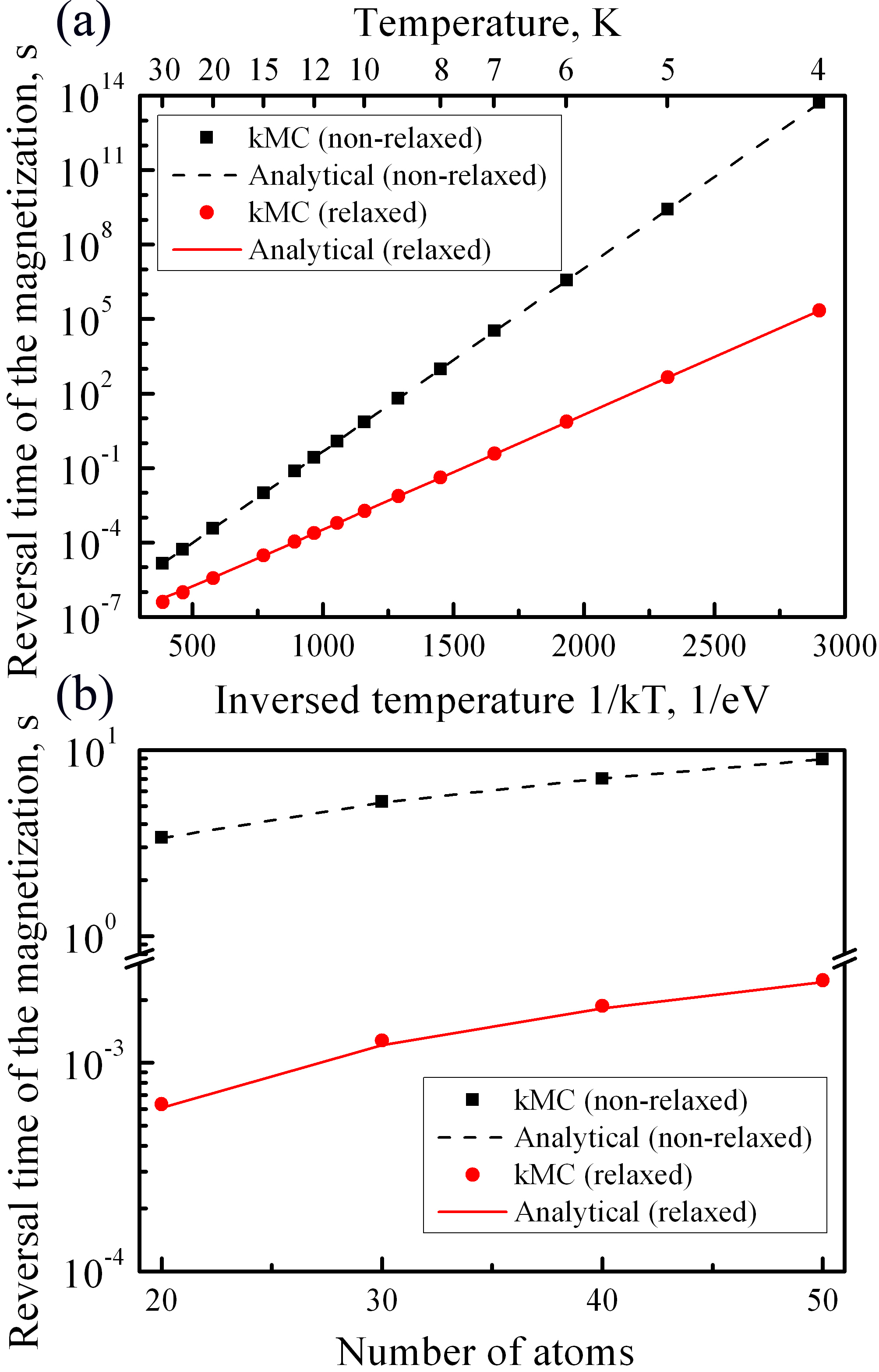}
\caption{\label{fig4} The dependencies of the reversal time of the magnetization the Co chain on Pt(997) surface in the framework the simple kMC model~\cite{PhysRevB.73.174418} and the improved kMC model~II. (a) The temperature dependence of the chain consisting of $N=40$ atoms. (b) The dependence on the length of the chain at $T=10$~K. The parameters of the Heisenberg Hamiltonian are the following: $J=7.5$~meV and $K=2.0$~meV.
}
\end{center}
\end{figure}

Figure~\ref{fig4} shows the dependencies of the reversal time of the magnetization the Co chain on Pt(997) surface. The results of kMC simulation are averaged over 1000 remagnetisations and shown with black (the simple kMC model) and red (the improved kMC model II) points. One can see that the relaxation effect leads to drastic decrease of the reversal time of the magnetization by several orders of magnitude at $T=4-30$~K (Fig~\ref{fig4}a). In order to understand this effect let as consider the Co chain consisting of $N=20$ atoms. The diffusion barriers for formation and disappearance of the domain wall at the edge of the chain are $E^D_1=10.7$~meV and $E^D_2=3.4\cdot10^{-3}$~meV. In the framework of the simple kMC model these events do not have diffusion barriers, and the equation (\ref{eqRate2}) gives a rude estimation of the transition rates: $\nu_1\approx\nu_0\exp(-2J/k_\text{B}T)$ and $\nu_2\approx\nu_0$. Comparing the values of $E^D_1$ and $2J=15$~meV we conclude that the relaxation effect leads to drastic increase of the rate $\nu_1$. The rate $\nu_3$ also dramatically increases. Indeed, the diffusion barrier for transitions of the domain wall along the chain (calculated in the middle of the chain) is $E^D_3=6.5\cdot10^{-3}$~meV. At the same time in the framework of the simple kMC model this barrier has a value of $K=2.0$~meV.

The reversal time of the magnetization the Co chains linearly increases with the increase of their length (Fig~\ref{fig4}b). It is clearly seen form the equation (\ref{eqTau2}) in the case of the simple kMC model. In the frame of the improved kMC model II the following relations take place at $T=4-30$~K: $\nu_1\ll \nu_2\approx\nu_3\approx\nu_0$. In this case the equation (\ref{eqTau}) can be simplified as $\tau\approx N_\text{eff}/4\nu_1$. It is interesting to note that the factor $\eta=\tau/\tau_\text{relaxed}$ slightly depends on the length of the Co chain as $\eta\sim(N-2)/(N-10)$.

The analytical results obtained with equation (\ref{eqTau}) are presented in Fig.~\ref{fig4} with solid and dashed lines. In the case of the simple kMC model the equations (\ref{eqRate1}) and (\ref{eqRate2}) are used for calculation of the transition rates $\nu_i$, $i=1,2,3$. In the case of the improved kMC model II we use the equation (\ref{eqRate3}) with the diffusion barriers $E^D_1$, $E^D_2$, and $E^D_3$ calculated for the Co chain consisting of $N=20$ atoms and presented above. Fig.~\ref{fig4} clearly shows that the analytical method gives the same results as the kMC simulations.

\section{Conclusion}\label{Conclusion}

Summarizing the results presented above we conclude that at all parameters of the Heisenberg Hamiltonian (\ref{eqHamiltonian}) the relaxation of the magnetic moments leads to the decrease of diffusion barriers and, consequently, to the decrease of the reversal time of the magnetization. If $J\ll K$ then the relaxation effect does not influence on the possible number of metastable states of the chain. The example of such system is an antiferromagnetic Fe chain on $\text{Cu}_2\text{N/Cu(001)}$ surface. The reversal time of the magnetization of this chain decreases by factor 2-3 at $T=4-7$~K. If $J\gtrsim K$, then the relaxation effect leads to the decrease of the effective length of the chain and to the appearance of clockwise and anti-clockwise states. The example of such system is the ferromagnetic Co chain on Pt(997) surface. The reversal time of the magnetization of this chain decreases dramatically by several orders of magnitude at $T=4-30$~K.

The presented improved kMC models can be easily generalized on the case of nonzero external magnetic field or on the case of interaction of the atomic chain with a STM tip. It is also very important to underline that the analytical approach~\cite{Kolesnikov1,Kolesnikov2,Kolesnikov3,Kolesnikov4} can be successfully used to estimation of the reversal time of the magnetization. The analytical method is incomparably less time-consuming than the kMC simulations, especially in the case of the improved kMC model II. A single domain-wall approximation is valid in a wide range of temperatures from the very low quantum tunneling temperature~\cite{JPCM27.455301} to the some maximal temperature which close to the critical temperature ($T_\text{max}\lesssim T_\text{C}$)~\cite{Kolesnikov1}. Thus, the analytical method can be a power tool for analyzing of magnetic properties of a wide class of atomic chains.

\section*{Acknowledgements}

The research is carried out using the equipment of the shared research facilities of HPC computing resources at Lomonosov Moscow State University~\cite{NIVC1,NIVC2}. The investigation is supported by the Russian Science Foundation (Project No. 21-72-20034).

\section*{Appendix. GNEB method in XY-model}

Let the atoms placed along the $y$-axis, $x$-axis is the easy axis of magnetization. The spherical angles $\theta_i$ and $\phi_i$ can be defined the in the following way ($s_i=1$): $(s_i)_x=\cos\theta_i$, $(s_i)_y=\sin\theta_i\cos\phi_i$, $(s_i)_z=\sin\theta_i\sin\phi_i$. According to the Heisenberg Hamiltonian (\ref{eqHamiltonian}) the magnetic energy of the chain can by written as
\begin{multline}\label{eqEnergy}
E=-\sum_{i>j}J_{ij}\left(\cos\theta_i\cos\theta_j+\sin\theta_i\sin\theta_j\cos(\phi_i-\phi_j)\right)\\
-K\sum_{i}\left(\cos\theta_i\right)^2.
\end{multline}
The necessary conditions of the local extremum or the saddle point is $\partial E/\partial\phi_i=0$, $i=1,\dots,N$.
At the arbitrary $\theta_i$ these equations have the solution $\phi_i=\phi_0$, where $\phi_0$ is some constant value. The energy (\ref{eqEnergy}) does not depend on $\phi_0$. Thus, we can choose $\phi_0=0$. It is so called XY-model.

In the most general case the state of a magnetic system consisting of N magnetic moments specified by 3N parameters. Taking into account the conditions $s_i=1$, $i=1,\dots,N$ one can choose a 2N-dimensional Riemannian manifold instead of 3N-dimensional Euclidean space~\cite{GNEB}. In the framework of XY-model the 2N-dimensional Riemannian manifold reduces to 1N-dimensional Riemannian manifold. The following description of the molecular statics and GNEB methods is fully analogous to Ref.~\cite{GNEB}. However, instead of singular equations (F.3-F.6) (see~\cite{GNEB}) we have very simple equations of a point mass motion on the 1N-dimensional Riemannian manifold
\begin{equation}\label{eqEM}
\frac{\text{d}v_i^\theta}{\text{d}t}=\frac{f_i^\theta}{m},~~~~~\frac{\text{d}\theta_i}{\text{d}t}=v_i^\theta,
\end{equation}
where $f_i^\theta$ and $v_i^\theta$ are projections of the $i$th force and the $i$th velocity on the orthogonal unit vector ${\bf e}_i^\theta$ in the direction of increasing $\theta_i$, and $m$ is effective mass. As a result, the searching of metastable states and calculations of diffusion barriers in the framework of XY-model are much more faster than in the general case. This is very important in the case of ``on the fly'' calculations.

\end{document}